\documentclass[fleqn,12pt,twoside]{article}
\usepackage{espcrc1}
\usepackage{epsfig}
\long\def\@makefigurecaption#1#2{#1. #2\par}

\newcommand{\AmS}{{\protect\the\textfont2
  A\kern-.1667em\lower.5ex\hbox{M}\kern-.125emS}}
\title{Strangeness Production in Proton--Proton Collisions Close to 
Threshold\thanks{Partly supported by the European Community -- Access 
to Research Infrastructure action of the Improving Human Potential 
Programme.}}
\author{M.~Wolke\address[fz]{IKP \& ZEL, Forschungszentrum J\"ulich, 
D--52425 J\"{u}lich, Germany}\thanks{present address: The Svedberg 
Laboratory, Thunbergsv\"agen 5A, Box 533, S--75121 Uppsala, Sweden.},
H.--H.~Adam\address[ms]{IKP, Westf\"alische Wilhelms--Universit\"at, 
D--48149 M\"unster, Germany},
A.~Budzanowski\address[ifj]{Institute of Nuclear Physics, PL--31--342 
Cracow, Poland},
R.~Czy\.zykiewicz\address[uj]{Institute of Physics, Jagellonian 
University, PL--30--059 Cracow, Poland}, 
D.~Grzonka\addressmark[fz], M.~Janusz\addressmark[uj], 
L.~Jarczyk\addressmark[uj], B.~Kamys\addressmark[uj], 
A.~Khoukaz\addressmark[ms], K.~Kilian\addressmark[fz], 
P.~Kowina\addressmark[fz], T.~Lister\addressmark[ms], 
P.~Moskal\addressmark[fz]\addressmark[uj]\thanks{talk given by 
P. Moskal.}, W.~Oelert\addressmark[fz], 
T.~Ro\.zek\addressmark[fz]\address[us]{Institute of Physics, 
University of Silesia, PL--40--007 Katowice, Poland}, 
R.~Santo\addressmark[ms], G.~Schepers\addressmark[fz],
T.~Sefzick\addressmark[fz], M.~Siemaszko\addressmark[us],
J.~Smyrski\addressmark[uj], S.~Steltenkamp\addressmark[ms],
A.~Strza{\l}kowski\addressmark[uj], P.~Winter\addressmark[fz],
P.~W\"ustner\addressmark[fz], W.~Zipper\addressmark[us]}

\begin{document}
\maketitle

\begin{abstract}
Exclusive data on the reactions $pp \rightarrow pp K^+ K^-$ and 
$pp \rightarrow p K^+ \Lambda/\Sigma^0$ have been taken at the 
cooler synchrotron COSY close to threshold.

At equal excess energies, an enhancement of the $\Lambda/\Sigma^0$ 
ratio by one order of magnitude has been observed compared to data at 
higher excess energies.
New results obtained at the COSY--11 facility explore the transition 
region between this low--energy $\Sigma^0$ suppression and excess 
energies of $60\,\mbox{MeV}$.

A first total cross section for elementary antikaon production 
below the $\phi$ threshold has been determined, two orders of 
magnitude smaller compared to kaon production at the same excess 
energy. 
\end{abstract}

\section{ELEMENTARY ANTIKAON PRODUCTION}
Studies on the reaction $pp \rightarrow pp K^+ K^-$ close to 
threshold have been motivated by the continuing discussion on the 
nature of the scalar resonances $f_0(980)$ and 
$a_0(980)$~\cite{Mor93+}.
Within the J\"ulich meson exchange model the $K \overline{K}$ 
interaction gives rise to a bound state in the isoscalar sector 
identified with the $f_0(980)$~\cite{Kreh97}.
Both shape and absolute scale of $\pi \pi \rightarrow K \overline{K}$ 
transitions crucially depend on the strength of the $K \overline{K}$ 
interaction, which in turn is a prerequisite of a $K \overline{K}$ 
molecule interpretation of the $f_0(980)$.
Similar effects might be expected in proton--proton scattering, and 
first results of exploratory microscopic calculations have recently 
been presented~\cite{Hai01b}.

\begin{figure}[hbt]
\parbox{0.42\textwidth}
  {\epsfig{file=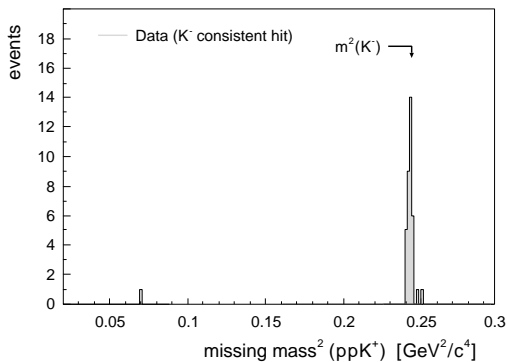,width=0.42\textwidth}} \hfill
\parbox{0.5\textwidth}
  {\vskip -10mm
   \caption{\label{fig:kkresult} Missing mass squared with respect to 
   an identified $(pp K^+)$ subsystem $17\,\mbox{MeV}$ above the 
   $K^+ K^-$ threshold. 
   The bin width corresponds to an experimental resolution of
   $\approx 2\,\mbox{MeV}/\mbox{c}^2$ (FWHM).}}
\end{figure}

A first total cross section value for the elementary antikaon 
production below the $\phi$ threshold in proton--proton scattering has 
been extracted from exclusive data taken at the COSY--11 installation 
at an excess energy of $\mbox{Q} = 17\,\mbox{MeV}$ with $\sigma = 
1.80 \pm 0.27^{+0.28}_{-0.35}\,\mbox{nb}$ including statistical and 
systematical errors, respectively~\cite{Quen01}.
The experimental technique is based on the measurement of the complete 
four--momenta of positively charged ejectiles.
Requiring furthermore a $K^-$ consistent hit in the dedicated 
negative particle detection system of the COSY--11 
facility~\cite{Brau96}, the identification of the four particle final 
state becomes (almost) completely free of background 
(fig.~\ref{fig:kkresult}).

\vskip -4ex
\vskip -4ex

However, the presently available statistics of $K^+ K^-$ events is not 
sufficient to distinguish non--resonant $K^+ K^-$ production and 
resonant production via the scalar resonances $f_0(980)$ and 
$a_0(980)$ from differential observables, e.g.\ the $K \overline{K}$ 
invariant mass distribution.

Considering the energy dependence of the total cross section, $\eta$, 
$\omega$ and $\eta^\prime$ production indicate strong imprints of 
final state interaction (FSI) at excess energies $\mbox{Q} \le 
100\,\mbox{MeV}$ in the proton--proton and, in case of $\eta$, in the 
proton--meson subsystems.
Contrary to this, $pp \rightarrow pp K^+ K^-$ cross section data 
obtained at COSY--11~\cite{Quen01} and DISTO~\cite{Bale01} below and 
above the $\phi$ threshold, respectively, are in reasonable agreement 
with one--boson exchange calculations~\cite{Sibi97} neglecting FSI 
effects.
Presently it is not clear whether the absence of the FSI influence in 
the $pp \rightarrow pp K^+ K^-$ reaction might be explained by a 
partial compensation of the $pp$ and $K^- p$ interaction in the final 
state or by the additional degree of freedom given by the four--body 
final state.
In the latter case FSI effects are expected to be more pronounced at 
energies very close to the $K^+ K^-$ production 
threshold~\cite{Sibi01}.

Data taking at excess energies closer to threshold and slightly below 
the $\Phi$ production threshold, i.e.\ at excess energies of 
$10\,\mbox{MeV}$ and $28\,\mbox{MeV}$ with respect to the $K^+ K^-$ 
threshold, has been successfully completed early this year at the 
COSY--11 facility and data analysis is presently in progress.

\section{EXCLUSIVE KAON--HYPERON FINAL STATES}
The most striking feature of the exclusive close--to--threshold data 
on $\Lambda$ and $\Sigma^0$ production in proton--proton scattering 
taken at the COSY--11 facility~\cite{Bale98,Sewe99} is the $\Sigma^0$ 
suppression with $\sigma\left(pp \rightarrow p K^+ \Lambda \right) / 
\sigma\left(pp \rightarrow p K^+ \Sigma^0 \right) = 28^{+6}_{-9}$ 
observed at equal excess energies below $\mbox{Q} = 13\,\mbox{MeV}$.
At excess energies $\ge 300\,\mbox{MeV}$ this ratio is known to 
be about 2.5~\cite{Bald88}.

Inclusive $K^+$ production data in $pp$ scattering from the SPES~4 
facility at an excess energy of $252\,\mbox{MeV}$ 
with respect to the $p K^+ \Lambda$ threshold show enhancements at 
the $\Lambda p$ and $\Sigma N$ thresholds of similar 
magnitude~\cite{Sieb94}.
Qualitatively, a strong $\Sigma N \rightarrow \Lambda p$ final state 
conversion might account for both the inclusive SATURNE results as 
well as the $\Sigma^0$ depletion in the COSY--11 data.
Evidence for such conversion effects is known e.g.\ from exclusive 
hyperon data via $K^- d \rightarrow \pi^- \Lambda p$~\cite{TTan69}.

\vskip -4.5ex
\begin{figure}[hbt]
\begin{center}
\makebox{\parbox{0.40\textwidth}
  {\epsfig{file=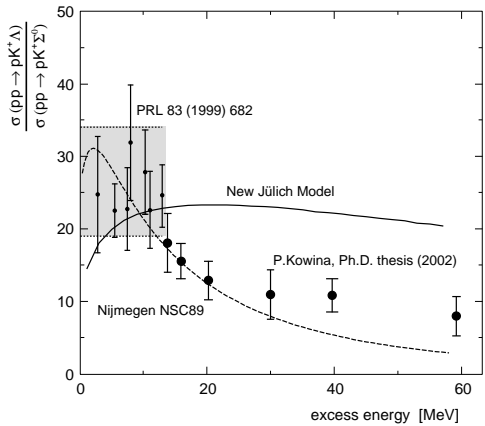,width=0.40\textwidth}}} \hfill
\makebox{\parbox{0.52\textwidth}
  {\vskip -10mm 
   \caption{\label{fig:lsratio} $\Lambda/\Sigma^0$ production ratio in 
   proton--proton collisions as a function of the excess energy. 
   Data within the shaded area are from~\cite{Sewe99}, results at 
   higher excess energies from~\cite{Kowi02}. 
   Calculations \cite{Gasp02} within the J\"ulich meson exchange 
   model assume a destructive interference of $K$ and $\pi$ exchange 
   and employ the microscopic $YN$ interaction models Nijmegen NSC89 
   (dashed line~\cite{Maes89}) and the new J\"ulich model 
   (solid line~\cite{Hai01a}), respectively.}}}
\end{center}
\end{figure}
\vskip -7ex

However, in exploratory calculations performed within the framework 
of the J\"ulich meson exchange model~\cite{Gasp00}, taking into 
account both $\pi$ and $K$ exchange diagrams in a coupled channel 
approach, a final state conversion is rather excluded as origin of 
the experimentally observed ratio.
While $\Lambda$ production is found to be dominated by kaon exchange 
both $\pi$ and $K$ exchange turn out to contribute to the $\Sigma^0$ 
channel with similar strength.
It is concluded~\cite{Gasp00}, that a destructive interference of 
$\pi$ and $K$ exchange might explain the close--to--threshold 
$\Sigma^0$ suppression.

An experimental study of $\Sigma$ production in different isospin 
configurations should provide a crucial test for the above 
interpretation, as for the reaction $pp \rightarrow n K^+ \Sigma^+$ 
an opposite interference pattern is found as compared to the 
$p K^+ \Sigma^0$ channel.
Measurements close to threshold are planned at the COSY--11 facility 
in the near future.

Contributions from direct production as well as heavy meson exchanges 
have been neglected so far in these calculations~\cite{Gasp00} but 
might influence the $\Lambda/\Sigma^0$ production 
ratio~\cite{Kais99}.
For complementary theoretical studies --- considering strangeness 
production close to threshold to proceed by one--boson exchanges or 
one--boson exchange followed by the excitation of nucleon 
resonances --- we refer to refs.~\cite{Tsu97+} and a recent 
review~\cite{Mosk02}.

Measurements on the $\Lambda/\Sigma^0$ production ratio in 
proton--proton collisions have been extended up to excess energies of 
$\mbox{Q} = 60\,\mbox{MeV}$ at the COSY--11 installation~\cite{Kowi02}.
In comparison to the experimental data, in figure~\ref{fig:lsratio} 
calculations are included obtained within the approach 
of~\cite{Gasp00} assuming a destructive interference of $\pi$ and $K$ 
exchange with different choices of the microscopic hyperon nucleon 
model to describe the interaction in the final state~\cite{Gasp02}.
The result depends on the details --- especially the off--shell 
properties --- of the hyperon--nucleon interaction employed.
At the present stage both the good agreement found in~\cite{Gasp00} 
for J\"ulich model~A~\cite{Holz89} with the close--to--threshold 
result and for the Nijmegen model (dashed line in 
fig.~\ref{fig:lsratio}) with the energy dependence of the cross 
section ratio should rather be regarded as accidental\footnote{In 
the latter case an SU(2) breaking in the ${}^3\mbox{S}_1$ 
$\Sigma N$ channel had to be introduced~\cite{Maes89} resulting in 
an ambiguity for the $\Sigma^0 p$ amplitude~\cite{Haid02}.}.
Calculations using the new J\"ulich model (solid line in 
fig.~\ref{fig:lsratio}~\cite{Hai01a}) do not reproduce the tendency of 
the experimental data.
It is suggested in~\cite{Gasp02} that neglecting the energy dependence 
of the elementary amplitudes and higher partial waves might no longer 
be justified beyond excess energies of $20\,\mbox{MeV}$.

The energy dependence of the total cross section for $\Lambda$ 
production up to excess energies of $\mbox{Q} = 60\,\mbox{MeV}$ is 
much better described by a phase space behaviour modified by the 
$p \Lambda$ final state interaction than by pure phase 
space~\cite{Kowi02}.
However, unlike the findings of~\cite{Sewe99} based on data up to 
$\mbox{Q} = 13\,\mbox{MeV}$, in the energy range up to 
$60\,\mbox{MeV}$ $\Sigma^0$ production is equally well described 
neglecting any FSI effect.
One reason for this qualitatively different behaviour might be, that 
the $\Sigma^0 p$ FSI is much weaker compared to the $\Lambda$--proton 
system.
On the other hand, a fit to the energy dependence based on a phase 
space behaviour implies dominant S--wave production and energy 
independent reaction dynamics as discussed above. 
Within the statistics of the present experiment, P--wave 
contributions can be neither ruled out nor confirmed at higher excess 
energies for $\Sigma^0$ production.
Consequently, high statistics $\Sigma^0$ data would be needed in 
future to study the onset of different partial waves experimentally.

\end{document}